\documentclass[twocolumn,showpacs,preprintnumbers,amsmath,amssymb,aps,floatfix,nofootinbib]{revtex4}

\usepackage{microtype}
\usepackage{hyperref}

\hypersetup{%
colorlinks=true,citecolor=blue,urlcolor=blue,linkcolor=blue,
  bookmarksnumbered=true,
  pdftitle = {},
  pdfsubject = {},
  pdfauthor = {A.~Alexandru},
  pdfkeywords = {}
}

\usepackage{graphicx}

\def\LL{\left\langle}	
\def\RR{\right\rangle}	

\newcommand{\BE}{\begin{displaymath}}
\newcommand{\EE}{\end{displaymath}}
\newcommand{\BNE}{\begin{equation}}
\newcommand{\ENE}{\end{equation}}
\newcommand{\BEA}{\begin{eqnarray}}
\newcommand{\EEA}{\nonumber\end{eqnarray}}

\newcommand{\Tr}{{\rm Tr\,}}

\def\fm {\,{\rm fm}}
\def\MeV {\,{\rm MeV}}

\def\chiPT{$\chi$PT}

\def\Tr {\mathop{\hbox{Tr}}}
\def\DU  {\mathop{{\cal D}\hbox{U}}}

\usepackage{dsfont}  
\usepackage{bm}      

\def\beq{\begin{equation}}
\def\eeq{\end{equation}}
\def\beqs#1\eeqs{\beq\begin{split} #1 \end{split}\eeq}

\def\pd#1#2{\frac{\partial #1}{\partial #2}}

\long\def\comment#1{}

\def\av#1{ \left\langle #1 \right\rangle }

\usepackage{multirow}
\usepackage{booktabs}

\begin{document}

\title{Sea contributions to the electric polarizability of the hadrons}

\author{Walter Freeman}\affiliation{The George Washington University}
\author{Andrei Alexandru}\affiliation{The George Washington University}
\author{Mike Lujan}\affiliation{The George Washington University}
\author{Frank X. Lee}\affiliation{The George Washington University}

\date{\today}

\begin{abstract}
We present a lattice QCD calculation of the polarizability of the neutron and 
other neutral hadrons that includes the effects of the background field on the sea quarks. 
This is done by perturbatively reweighting the charges of the sea quarks to couple 
them to the background field. The main challenge in such a calculation is stochastic 
estimation of the weight factors, and we discuss the difficulties in this estimation. 
Here we use an extremely aggressive dilution scheme to reduce the stochastic noise to a 
manageable level. The pion mass in our calculation is $300\MeV$ and the lattice size
is $3\fm$. For neutron, we find that $\alpha_E = 2.70(55)\times 10^{-4}\fm^3$, which is
the most precise lattice QCD determination of the polarizability to date that includes sea effects.
\end{abstract}

\maketitle

\section{Introduction}

At leading order, the interaction of hadrons with a background electromagnetic 
field can be parametrized with a variety of electromagnetic polarizabilities which 
characterize the deformation of the hadron by the field. Of these, the electric 
polarizability $\alpha$ describes the induced dipole by an external static, uniform 
electric field. It is defined as the ratio of the electric field and the induced dipole 
moment: $\bm d = \alpha\,\bm{\mathcal E}$. 
Since this deformation is a direct consequence of the composite nature of the hadrons, 
it is a necessary component of any overall understanding of hadronic structure.

Measuring the electric polarizabilities for hadrons is challenging. Few hadron
polarizabilities have been determined, but there are a number of experiments that plan 
to measure these quantities for various hadrons in the near future. On the theory
side lattice QCD can be used to determine these parameters as predicted by quark-gluon
dynamics. These are challenging calculations, and to establish the methodology it is
useful to first focus on electrically neutral hadrons, which are not accelerated by
the electric field. Since the hadrons are at rest, it is easier to detect the effect 
of electric polarizability. In this paper we focus on the neutron but we will also
present results for neutral kaon and neutral pion.

Neutron electric polarizability is difficult to measure experimentally, due to the 
unavailability of free neutron targets. It has been measured in the laboratory by 
scattering neutron beams on lead~\cite{Schmiedmayer:1991zz} and off of 
deuterons~\cite{Kossert:2002ws}; the results respectively were 
$12.0(1.5)(2.0) \times 10^{-4}\fm^3$ and $12.5(1.8){+1.1 \choose -0.6} \times 10^{-4}\fm^3$. 

A lattice calculation of the neutron electric polarizability is desirable for at 
least three reasons. First, the experimental uncertainties in these 
quantities are still over $10\%$, and it may be the case that eventually the lattice 
may prove superior to experiment in attaining a precision measurement of this quantity. 
Second, if lattice QCD is to be considered a successful approach to simulating the 
hadronization of quarks and their properties, then the measurement of such a fundamental 
property of the neutron is something of a basic test.  Finally, the flexibility of 
lattice calculations (the freedom to use nonphysical parameters) may provide some insight 
into the origins of the neutron polarizability. 

The first lattice study of the neutron polarizability was done in 1989~\cite{Fiebig:1988en}, 
on a $10^3 \times 20$ quenched lattice with $a \simeq 0.11\fm$ using unimproved 
staggered fermions; this study, along with a subsequent early study using both Wilson and 
clover fermions on a quenched sea~\cite{Christensen:2004ca}, show good agreement with the 
experimental value. More recently, improved calculations have produced values that 
are substantially smaller~\cite{Engelhardt:2007ub,Alexandru:2008sj,Alexandru:2009id,
Detmold:2008xk,Detmold:2009dx,Detmold:2010ts}, suggesting that the early agreement 
with experiment was coincidental. 

It is well understood that neutron polarizability computed from lattice QCD is smaller 
than the physical value because the quark mass used is heavier than the physical
one. Chiral perturbation theory (\chiPT) predicts that the polarizability of the
neutron diverges in the chiral limit. In fact \chiPT\ calculations can be used to
predict the value of the polarizability for unphysical quark 
masses~\cite{Hildebrandt:2003fm,McGovern:2012ew,Lensky:2009uv}. The most precise lattice
QCD calculation for the neutron polarizability finds a value that is still 
significantly different from the \chiPT\ predictions~\cite{Lujan:2014kia}. 
The difference is most likely due to a combination of finite volume effects and 
a systematic correction due to the electric charge of the sea quarks.
In this paper we present a method for removing the latter systematic error and use
it to compute correct value of the polarizability on one of the ensemble used 
in our previous study~\cite{Lujan:2014kia}.

Since lattice QCD is best able to measure spectroscopic
information about hadronic states, we compute the polarizability 
through the induced interaction energy $\delta E = -\frac{1}{2} \alpha {\cal E}^2$.
This is achieved using the {\em background field method}, where the energy 
shift is computed by comparing the mass of the hadrons in the presence
of a static electric field with the one determined when the field is absent.
To include the effects of the charged sea, one could generate two dynamical ensembles, 
one with a background field and one without, and measure the mass shift. 
However, for the valence-only calculation these two masses, measured on the same 
Monte Carlo ensemble and differing only by the effects of a perturbatively-small 
background field, are highly correlated, and thus the error on the mass shift is much less
than the error on each mass individually. Generating two separate ensembles would 
destroy this correlation and greatly inflate the statistical error.
What is needed is a way to obtain ensembles generated with different dynamical properties 
which are correlated; reweighting provides such a technique.

The plan of the paper is the following: in Section~\ref{sec:valence} we will review briefly
the steps relevant for the valence calculation. In Section~\ref{sec:reweighting} we discuss
the perturbative reweighting strategy we use to couple the sea quarks to the background field. 
In Section~\ref{sec:estimator} the stochastic estimators used to compute the derivatives
of the reweighting factors are discussed in detail. The results are presented in Section~\ref{sec:results}.

\section{Valence calculation}
\label{sec:valence}

\subsection{Simulation parameters}

In this study we will use one of the ensemble from a previous study~\cite{Lujan:2014kia},
labeled EN1 in that paper. The configurations in this ensemble were generated using
$N_f=2$ flavors of nHYP-smeared Wilson-clover fermions~\cite{Hasenfratz:2007rf}.
The ensemble contains 300 lattice configurations of
size $24^3 \times 48$. The lattice spacing of $0.1245(16)\fm$ was determined by a fit to 
the static quark potential to determine the Sommer scale $r_0/a$~\cite{Sommer:1993ce}
using a value of $r_0=0.5\fm$. 
The sea quarks have $\kappa=0.1282$, corresponding to $m_\pi=306(1)\MeV$; we use the
same $\kappa$ value for the valence light quarks as well. The valence strange quark 
for the kaon correlators has $\kappa_s=0.1266$. The gauge configuration
generation was performed with periodic boundary conditions;  
Dirichlet boundary conditions have been applied for the valence quarks in the direction
of the electric field and the time direction. We use an optimized multi-GPU Dslash operator~\cite{Alexandru:2011fh} 
along with an even-odd preconditioned BiCGstab inverter~\cite{Alexandru:2012ja} to do the analysis described here.

\subsection{The background field method}

Since the ground state energy of the neutron is shifted by an amount 
$\delta M = -\frac{1}{2} \alpha \mathcal{E}^2$ in an external electric field, spectroscopic 
measurements on the lattice can provide a direct avenue to access the polarizability.
We use the notation $\delta M$ rather than $\delta E$ to emphasize that, 
since we use Dirichlet boundary conditions, we do not measure the actual neutron 
mass since we have no zero-momentum state.
The approach is straightforward: we measure the neutron energy with the background 
field and without it, then compute $\delta E$, which is then converted to $\delta M$ 
to compute $\alpha$.

We introduce the electric field by adding a $U(1)$ phase factor on top of the 
$SU(3)$ gauge links that corresponds to a uniform background field; this may be done 
in any convenient choice of gauge. In practice, there are several complications which 
must be taken into account when applying this method to the lattice. The simplest is 
the fact that in Euclidean time, applying phase factors of the form $e^{i\theta}$ 
corresponds to an imaginary electric field; to get a real electric field, one must 
use an imaginary $\theta$, giving real exponential factors on the links. However, 
an imaginary electric field presents no real problems; this gives a positive 
$\delta E$ as expected, and has little effect on the final result~\cite{Alexandru:2008sj}.

We also must address the lattice boundary conditions. With periodic boundary 
conditions, the phase factor corresponding to an arbitrary electric field will 
have a discontinuity at the lattice edge, giving a non-physical spike in the electric 
field there. While we can choose values of $\bm{\mathcal E}$ in conjunction
with the lattice size and gauge such that the discontinuity is made to vanish, 
the size of $\bm{\mathcal E}$ required to do this is so large
that one is no longer probing only the lowest-order effects proportional to ${\mathcal E}^2$ 
for which the polarizability is defined. Moreover,
even if the discontinuity in the $U(1)$ phase is addressed, the electric scalar 
potential will not be single-valued, possibly inducing
quark lines or charged pions to wind repeatedly around the lattice. 
It is not clear what the effects of this, or of discontinuities
in the $U(1)$ phase, will be.

Thus, we choose to use Dirichlet boundary conditions in time and in the direction of 
the electric field, which we choose as the $\hat x$-direction. While this means that 
we have no true zero-momentum state, this can be treated as an additional
finite-size effect whose effect can be partially compensated for and which will in 
any case go away in the infinite-volume limit.

We parametrize the electric field with the dimensionless parameter
\beq
  \eta \equiv a^2 q {\mathcal E}\,,
\eeq
noting that $\eta$ depends both on the quark flavor and $\mathcal E$, and choose a 
gauge for the electric field such that
\beq
  U_4 \rightarrow U_4 e^{i \eta x/a}\,.
\eeq
$\eta_d$ must be chosen small enough that it probes only the lowest-order~(quadratic) 
effects which correspond to the electric polarizability and avoids large 
$\mathcal O(\mathcal E^4)$ effects. Since there are two sea quarks, we use a 
different $\eta$ for up and down quark propagators, and quote $\eta_d$ as a measure of 
the field strength. However, choosing a value which is too small means that we 
may encounter issues with numerical precision, either with the accuracy of inverters or 
(in the extreme case) machine precision.

\begin{figure}[!t]
  \centering\includegraphics[width=0.45\textwidth]{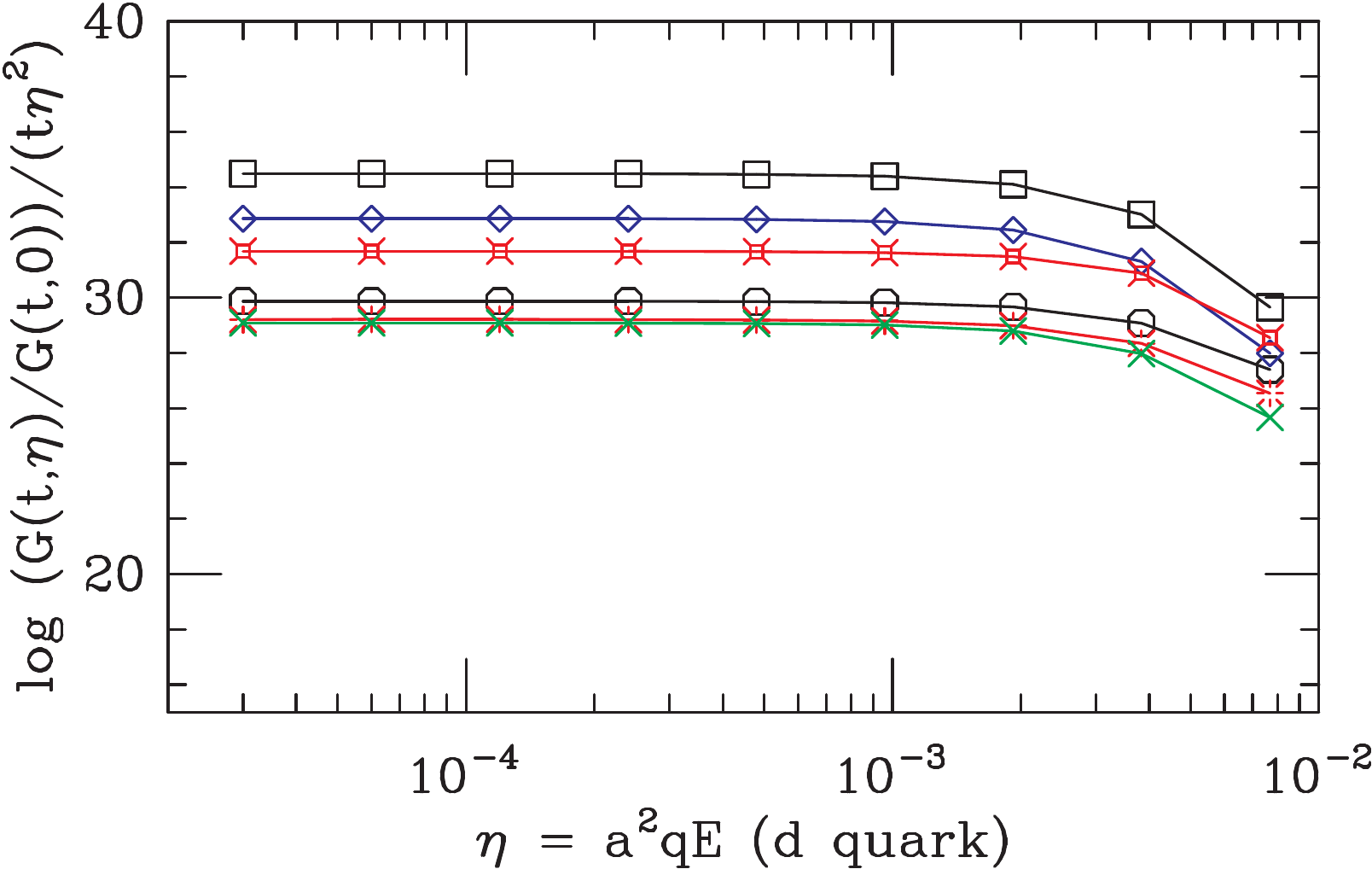}
  \caption{Dependence of $(\log \frac{G_\pi(t,\eta)}{G_\pi(t,0)})/(t \eta^2)$ for the 
  neutral pion on a $24^3 \times 48$ lattice,
    as a function of $\eta$ for different correlator times.
    This quantity is roughly equivalent to the shift in the
    effective mass divided by $\eta^2$,
    and should be constant in the range where $\eta$ creates a purely quadratic effect.}
  \label{fig-eta-scaling}
\end{figure}

Fig.~\ref{fig-eta-scaling} shows the response of the neutral pion correlator, $G_\pi$, to the 
background field as a function of $\eta_d$ for a few different correlator times on 
one configuration. The breakdown of quadratic scaling as $\eta$ becomes large is clear.
Note that $G_\pi(t,\eta)$ is symmetrized with respect to $\eta$, so that only even powers of
$\eta$ contribute to this correlator. This symmetrization is only valid when the sea 
quarks are not charged.
In this study we used $\eta_d = 10^{-4}$, and the valence correlators on this ensemble 
were run at this value. Fig.~\ref{fig-eta-scaling}
shows that this value is well within the quadratic scaling region, at least for the
valence contribution.

\subsection{Extracting the energy shift}
\label{sec-stats}

To determine the energy shift caused by the external electric field, we compute hadron 
correlators $G(t,{\mathcal E}_x)$ for positive, zero, and negative 
values of $\mathcal E_x$. This requires the computation of five quark-line propagators: 
one at $\eta=0$, used for both up and down quarks in the case of $\mathcal E_x=0$, 
two at $\eta=\pm \frac{1}{3} a^2 \mathcal E_x$ for the down quark, and two at 
$\eta=\mp \frac{2}{3} a^2 \mathcal E_x$ for the up quark.

The energy shift caused by the external electric field is quite small, smaller than the 
stochastic error in the hadron energy itself. Thus, in order to resolve it, we must take into account the 
fact that the correlators measured with and without the electric field are strongly 
correlated, and only become more strongly correlated as the strength of the electric 
field is decreased. We cannot simply, then, do independent correlator fits to the three
correlators. Just as an ordinary correlator fit must take into account the correlations 
between $G(t)$ at different $t$ by computing the covariance between them, we must 
construct a covariance matrix which includes the mixed covariance between zero-field 
and nonzero-field correlators. This is simply an extension of the standard fitting
procedure using the covariance between all pairs of observables.

We then fit all the data at once, using the fit form
\beq
  \av{G(t,\eta)}_\eta = (A + \Delta A) e^{-(E + \Delta E)t} \,,
\label{eq:model}
\eeq
to extract $E$ and the parameter $\Delta E$ which is related to $\alpha$. For details on 
determining the polarizability from the energy shift, see~\cite{Lujan:2014kia}.

For small values of $\mathcal E_x$, the covariance matrix is quite poorly conditioned 
due to the extremely strong correlations. We have observed that in this case both the
minimization of $\chi^2$ and the inversion of the covariance matrix must
be done in extended precision to get consistent fit results. 
For $\eta_d=10^{-4}$, we find that the C {\tt {long double}} type offers sufficient precision.

\section{Reweighting}
\label{sec:reweighting}

\subsection{General remarks}

As mentioned previously, the simplest way to incorporate the effect of the electric field 
on the sea quarks would be to include its effects in gauge generation where the sea 
dynamics are simulated. However, generating a separate Monte Carlo ensemble to compute 
the correlator in the presence of background field would ruin the correlations which 
are necessary to achieve a small overall error. Thus, we turn to reweighting as a method 
of creating two ensembles which have different sea-quark actions yet are correlated. A 
similar approach has been used before to compute the strangeness of the nucleon using 
the Feynman-Hellman theorem~\cite{Ohki:2009mt}, which requires a measurement of 
$\pd{M_N}{m_s}$.

Reweighting involves a simple modification of Monte Carlo sampling. Normally, 
the configurations are sampled using a probability proportional to $e^{-S}$. 
Then a Monte Carlo estimate for the expectation 
value of the correlator $G(t)$
\beq
  \av{G(t)}_0 \equiv \frac{\int \DU  G(t) {e^{-S_0}}}{\int \DU {e^{-S_0}}} 
  \approx \frac1{N_\text{confs}} \sum_{i=1}^{N_\text{confs}} {G(t)_i} \,.
\eeq
If we instead want to simulate the physics of a different action 
$S_{\eta}$~(in our case, with the background electric field)
but have access to Monte Carlo configurations using the action $S_0$, 
we can simply modify the Monte Carlo estimate
to correct for the additional portion of the factor $e^{-S}$:
\beq
  \av{G(t,\eta)}_{\eta} = 
  \frac{\av{G(t,\eta) e^{-(S_\eta-S_0)}}_0}{\av{e^{-(S_\eta-S_0)}}_0}
  \approx
\frac {\sum_i{G(t,\eta)_i w_i}}{\sum_i w_i} \,,
\label{eq:rew}
\eeq
where $\av{\cdot}_0$ indicates the average with respect to $e^{-S_0}$ and
$w_i \equiv e^{-(S_{\eta} - S_0)}$ is the reweighting factor 
associated with configuration $i$.

The contribution to the weight factor from 
the fermion sector, using the standard prescription where the fermions are integrated out, 
can be written as a ratio of fermion determinants:
\beq
w_i = \frac{\det M_{\mathcal E} (U_i)}{\det M_0 (U_i)} \,.
\eeq
We want to include the effect of the electric field on both flavors of sea quarks; this can be done by simply computing weight
factors at two values of $\eta$ (corresponding to the up and down quark charges) and multiplying them.

There are two well-known problems associated with reweighting. The first is that if 
the overlap between the target and simulated ensemble is poor,
the weight factor fluctuates too strongly and the reweighted ensemble will wind up 
dominated by just a few configurations, leading to a lack of statistical power. The second
is that the determinant ratio must be estimated stochastically. The good news is that the 
since the average over stochastic noises commutes with the gauge average, any unbiased 
estimator for the weight factor will also produce an unbiased estimate for operators 
computed on the reweighted ensemble~\cite{Hasenfratz:2008fg}, even if it is quite noisy. 

When the reweighting factors are close to one, the overlap is good and for most
estimators the stochastic noise is also reduced. Since we can get the reweighting
factors arbitrarily close to one by decreasing the value of $\eta$, none of the
issues mentioned above create problems for our calculation. On the other hand, this does not mean
that our calculation gets more precise as $\eta\to0$. This is because the signal
we try to measure is encoded in the correlation between the weight factor and
the ones in the hadronic correlator. As $\eta$ is decreased both signal and
error decrease in concert, leading to a constant relative error.

\subsection{Perturbative reweighting}
\label{semi-pert-reweighting}

As we will see, the most difficult part of performing the reweighting calculation 
for the electric field is the estimation of the weight factors, as the stochastic 
estimators for the weight factor in our case are substantially more noisy than 
in the traditional mass reweighting. 

Stochastic estimators for determinant ratios have been used in many studies, more
recently as a technique to fine-tune the quark mass in dynamical simulations
via reweighting~\cite{Ohki:2009mt,Hasenfratz:2008fg,Liu:2012gm}.
We attempted at first to use a similar method to estimate the weight factors. 
However, even for large numbers of stochastic noises, we were unable to resolve even 
the difference of the weight factors from unity on a production-sized 
lattice~\cite{Freeman:2012cy}.

In this study we use an alternative to the standard stochastic estimator, 
a perturbative technique for estimation 
of the weight factor. Since we are interested only in perturbatively small $\bf \mathcal E$, 
we can expand the one-flavor weight factor $w_q$ about $\eta=0$:
\beq
w_q(\eta) = 1 + \eta \left. \pd{w_q}{\eta} \right|_{\eta=0} + 
\frac12 \eta^2 \left. \pd{^2w_q}{\eta^2} \right|_{\eta=0} + \mathcal O(\eta^3) \,.
\label {eq-pert-weight}
\end{equation}
To obtain the two-flavor weight factor $w$ at some particular value of $\bf \mathcal E$
corresponding to $\eta_d$ for the down quark and $-2\eta_d$ for the up quark, 
we simply multiply, keeping terms only up to $\eta^2$:
\beqs
\label{eq-expansion}
w(\eta_d) &= w_d w_u = w_q(\eta_d) w_q(-2\eta_d) \\
  &= 1 - \eta_d \pd{w_q}{\eta} 
  +\eta_d^2 \left[\frac52  \pd{^2w_q}{\eta^2} -
  2 \left(\pd{w_q}{\eta}\right)^2 \right] \,.
\eeqs
The derivatives are computed for $\eta_d=0$. To simplify notation, we will
denote the derivatives with respect to $\eta$ around $0$ as $w_q'$ and $w_q''$.
Given estimates of these derivatives, 
we can evaluate the above at any sufficiently-small $\eta_d$ to produce 
a reweighted ensemble on which to apply the valence calculation. 
This is a {\it semi-perturbative} calculation, since the sea effects are introduced 
perturbatively {\it via} the perturbative estimates of the weight factors, but these weight 
factors are evaluated at finite $\eta_d$ and used as inputs to the valence calculation.
This differs from the full-perturbative method introduced by 
Engelhardt~\cite{Engelhardt:2007ub,Engelhardt:2010tm} in that the hadron correlators
are computed non-perturbatively, for a small value of $\eta$.
This allows one list of weight factors to be applied to a variety of hadrons, which would 
not be possible in a fully perturbative calculation. Since determination of the weight 
factors requires the majority of the computational effort, the numerical effort is
greatly reduced when computing the polarizability for a set of hadrons.

Note that we did not include a contribution from the strange quarks in the
perturbative expansion. In part this is because the strange sea quarks were
not included in the measure used to generate our gauge configurations. Additionally, 
to include the correction due to the electric charge of the sea strange
quarks requires evaluating the derivatives $w'_s$ and $w''_s$ for a different 
quark mass, significantly increasing the numerical effort, while their contribution 
is expected to be extremely small. 

While we are looking only for quadratic effects and expect no shift in the neutron mass 
proportional to $\eta_d$ (due to reflection symmetry), these can arise in two ways: either by 
the sole effect of the quadratic term in the weight factor, or by a correlation between 
the first-order term in the weight factor with a similar linear effect in the neutron 
correlator. The latter occurs because reflection symmetry is not preserved configuration 
by configuration, but only in the gauge average. We expect that the gauge average of 
$w_q'$ is zero, but on individual configurations it will be nonzero.

To evaluate the derivatives we can use Grassman integral techniques and we get
\beq
w_q' = \pd{}\eta \frac{\det M_\eta}{\det M_0}=
\Tr \left( M' M_0^{-1} \right) \,,
\label{eq:9}
\eeq
and
\beqs
w_q''=& \pd{^2}{\eta^2} \frac{\det M_\eta}{\det M_0} = 
\Tr \left( M'' M_0^{-1} \right) \\
&+
\left(\Tr M' M_0^{-1} \right)^2 - \Tr \left(M' M_0^{-1} \right)^2 \,,
\label{eq:10}
\eeqs
where $M'$ and $M''$ are the derivatives with respect to $\eta$ at $\eta=0$ 
of the one flavor fermionic matrix $M$.

\begin{figure}[!t]
  \centering\includegraphics[width=0.49\textwidth]{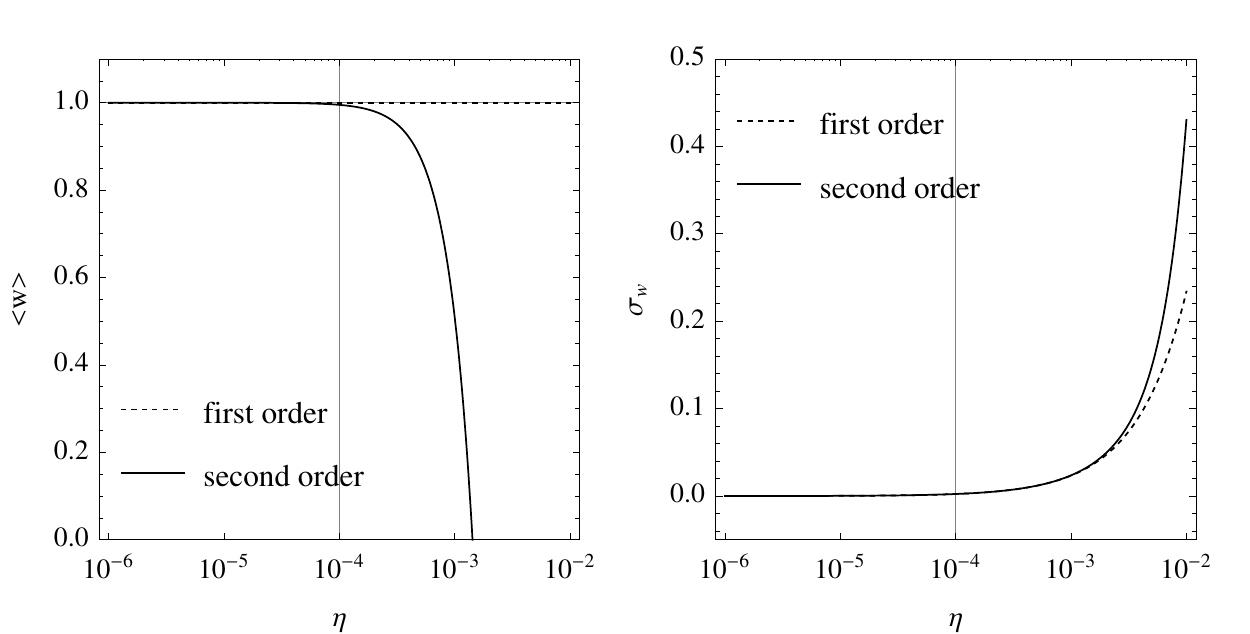}
  \caption{Dependence of the mean and standard deviation of the reweighting
  factor as a function of $\eta$ for the $24^3 \times 48$ ensemble used in
  this study. The estimator is computed using Eq.~\ref{eq-expansion}. The solid 
  lines include the quadratic effects in $\eta$, whereas
  the dotted lines include only the linear term. Vertical lines indicate
  the $\eta$ value we used.}
  \label{fig-w-vs-eta}
\end{figure}

Once the derivatives are computed, we can use them to determine the
values of $\eta$ that are in the small field region. Since we do a perturbative
expansion, we want to make sure that the higher order terms are not important.
It may seem that given that we only keep the terms of interest, we can set $\eta$
to any value -- the higher order terms are not present. On another hand, the
reweighting is successful only when $\av{w}$ is close to one. If we choose
values of $\eta$ that are too large, the individual reweighting factors could
even go negative. In fact, we can choose $\eta$ such that 
$\av{1+w'\eta +w''\eta^2/2}=0$.
It is unclear that the results of the reweighting are meaningful
in this case. To set bounds on the $\eta$ value we used as a guiding principle
the requirement that the Taylor expansion of $w(\eta)$ is a good approximation.

For the Taylor expansion to be successful, we expect that the successive terms
in the expansion are subdominant. We ask then that $\eta$ be such that 
$1\gg w'\eta \gg w'' \eta^2/2$. In Fig.~\ref{fig-w-vs-eta}
we show both the mean $\av{w}$ and standard deviation $\sigma_w$ for our ensemble
as a function of $\eta$ when using the first and second order approximations for $w$. 
The mean $\av{w}$ when including only the first order term is close
to one for all values of $\eta$ since $\av{w'}\approx 0$, as demanded by symmetries.
Note that the mean when including the quadratic term in the approximation 
deviates quickly from one as we increase $\eta$. This is due to the large value of $\av{w''}$.
In fact, a large constant $\av{w''}$ is not important since it cancels out in the
reweighting ratio from Eq.~\ref{eq:rew}. The fluctuations of $w$ about the mean are important and that is why we
plot $\sigma_w$ as a function of $\eta$. Note that the standard deviation is dominated
by the first order term for values of $\eta$ much larger than the ones where the mean
deviates from one.
In any case, the value of $\eta=10^{-4}$ used in this study is well inside the region
where the Taylor expansion is working well.

\section{Stochastic estimations of the weight factor}
\label{sec:estimator}

The traces that appear in the expressions for the determinant derivatives, 
Eqs.~\ref{eq:9} and~\ref{eq:10}, can be evaluated stochastically in the 
standard way, that is
\beq
\Tr \mathcal O = \av{\xi^\dagger \mathcal O \xi}_\xi \,,
\eeq
where $\xi$ are Z(4) noise vectors. We note that only three estimators are 
required---$\Tr M'M_0^{-1}$, $\Tr M''M_0^{-1}$, and $\Tr (M'M_0^{-1})^2$---since 
an estimator for $\left[ \Tr (M'M_0^{-1}) \right]^2$ can be constructed 
from two uncorrelated values of the estimator for the first-order term 
$\Tr (M'M_0^{-1})$. 
As $\left[ \Tr (M'M_0^{-1}) \right]^2$ is both computed separately
and subdominant, we refer to the combination of the two second-order terms that 
must be explicitly estimated, $\Tr (M''M_0^{-1}) - \Tr (M'M_0^{-1} M'M_0^{-1})$, 
as ${\tilde w}_q''$.
Note that there is no bias introduced by using the same stochastic noise vector 
for the $w_q'$ and $\tilde w_q''$, since the ultimate computation of the weight factor 
involves only linear combinations of these estimates; any correlations in the 
stochastic fluctuations will not cause the final result to be biased.
This reduces the number of inversions required per noise vector to two.

\begin{figure}
  \centering\includegraphics[width=0.45\textwidth]{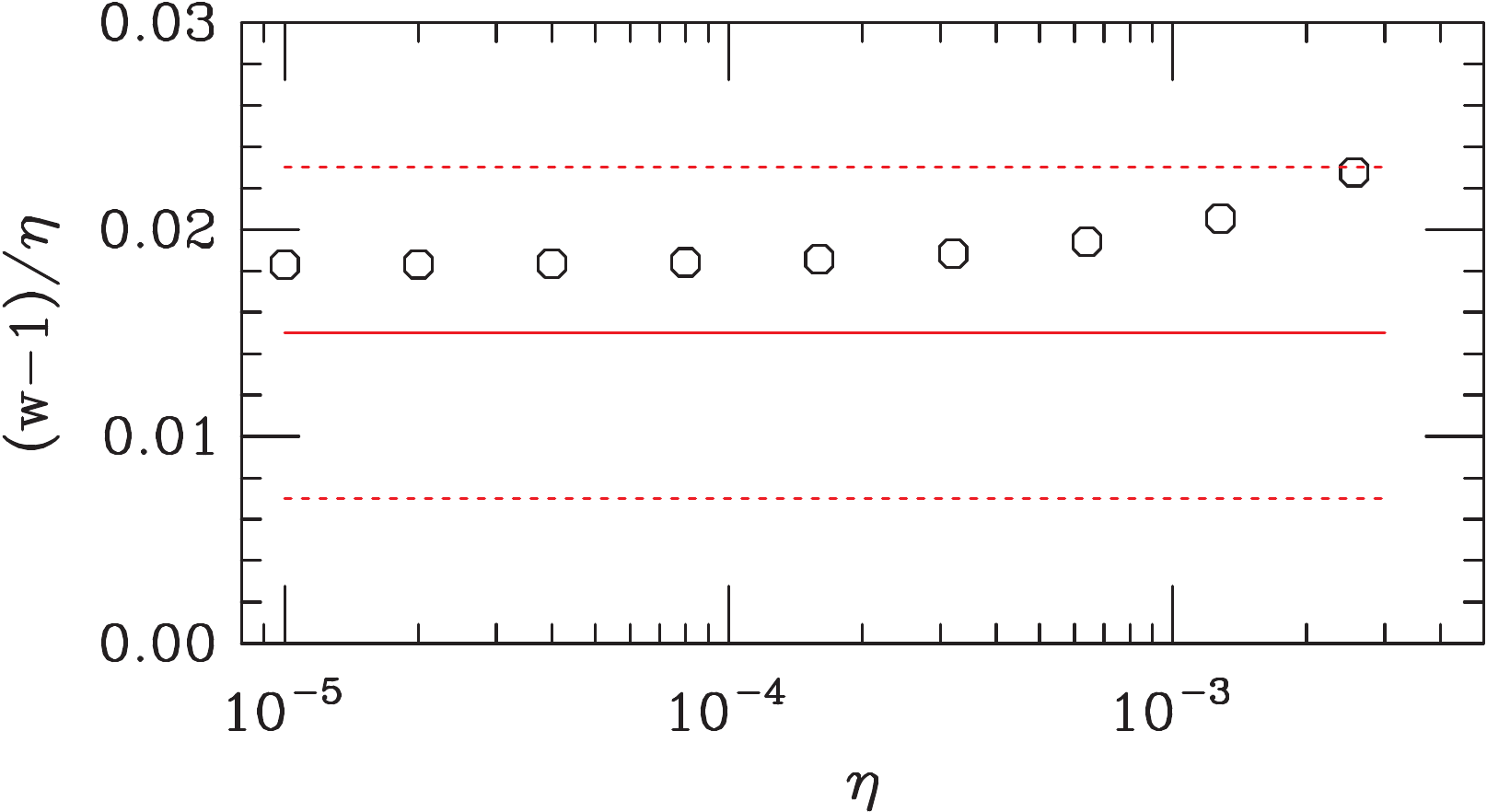}
  \caption{Exact values for $[w(\eta)-1]/\eta$ on a $4^4$ lattice, compared with the 
  value predicted by the stochastic estimator for $w'$ and its error band.}
  \label{deriv-confirm}
\end{figure}

Standard stochastic estimators of these traces are, unfortunately, very noisy. 
For example, on a $4^4$ lattice we need $5\times 10^6$ noise vectors 
to obtain a signal-to-noise ration greater than one for the first derivative. 
In Fig.~\ref{deriv-confirm}
we compare the stochastic result with the exact result computed via direct 
evaluation~\cite{Alexandru:2010yb}. We see an agreement between $w_q(\eta)$ 
and the stochastic estimator for $w_q'$, with the onset of quadratic behavior 
visible as $\eta$ is increased.

\begin{figure*}[t]
\begin{center}
\includegraphics[width=0.45\textwidth]{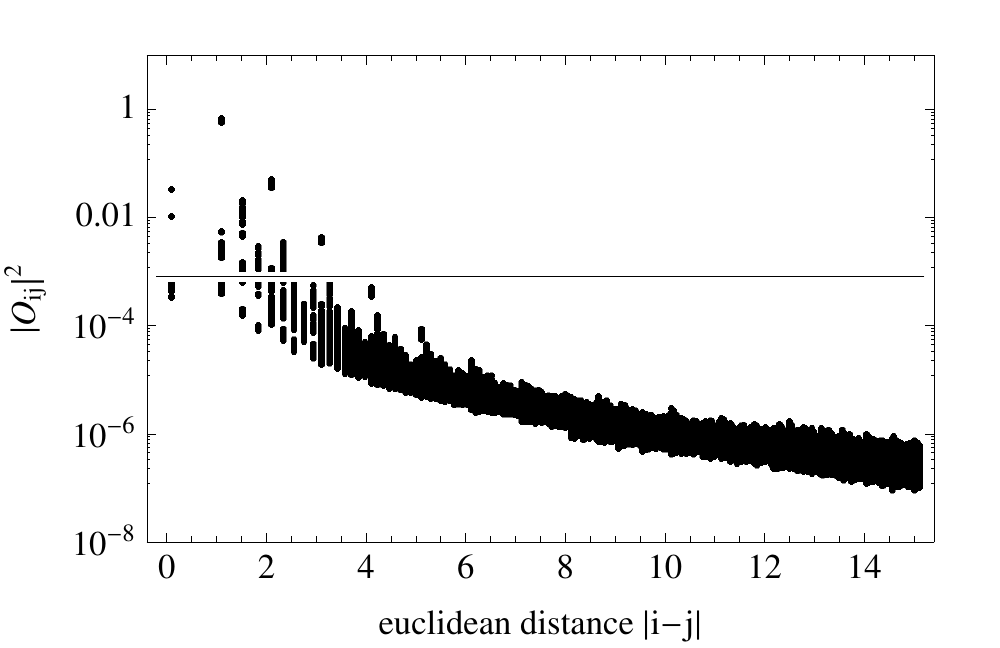}\kern1cm
\includegraphics[width=0.45\textwidth]{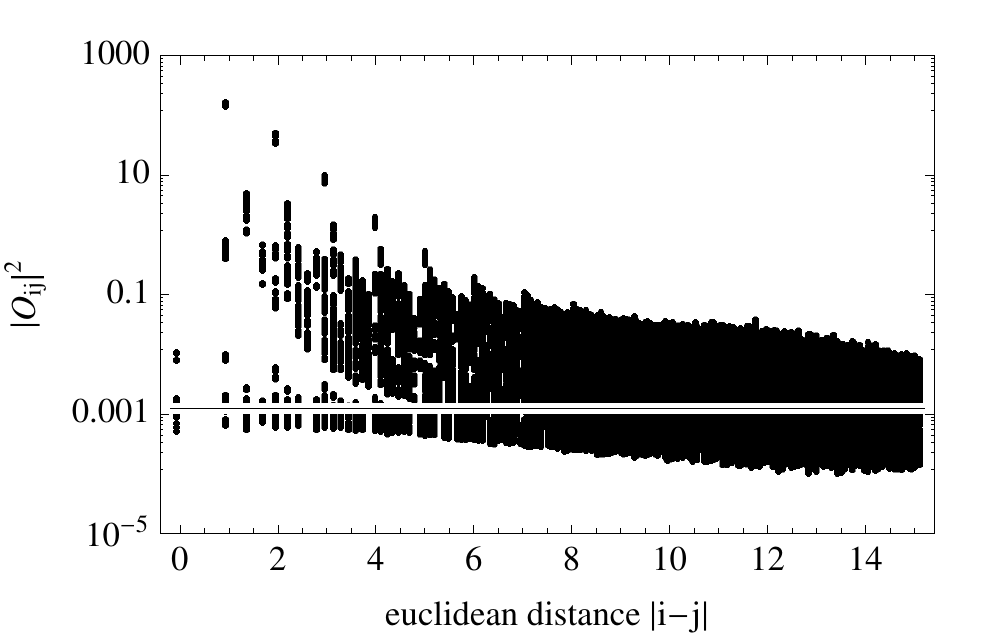}
\end{center}
\caption{Mapping of a representative set of the off-diagonal elements of 
$M' M_0^{-1}$ (left) and $M''M_0^{-1} - (M'M_0^{-1})^2$ (right). 
The average of the square of the $12$ diagonal elements, 
the ones that define the signal, is shown as a horizontal line. 
}
\label{fig-ope-map}
\end{figure*}

\subsection{Estimator quality}
\label{sec-est-qual}

Since the limiting factor for this calculation is the stochastic estimation 
of the weight factors, it is useful to understand how far we need to reduce 
the variance in the stochastic estimator. Whether using perturbative or 
nonperturbative reweighting, it is the variation of the weight factor between 
gauge configurations that carries the information, and it is this fluctuation 
that we seek to extract using a stochastic estimator. Thus the gauge variance 
between configurations in the weight factor amounts to a signal, while the 
stochastic variance gives the noise in that signal. This immediately suggests 
a criterion for judging the quality of any given stochastic estimation scheme: 
the stochastic signal-to-noise ratio
\begin{equation}
w_{\mathrm SNR} \equiv \frac{\sigma_{\rm gauge}}{\sigma_{\rm noise}} \,.
\label{SNR}
\end{equation}
Ideally we would like this SNR to be as large as possible. A SNR substantially 
less than unity means that the stochastic estimator scheme used is insufficient 
to extract whatever physics differences exist between the original and reweighting 
ensembles. In our case, this may be because the actual difference is small, or 
because the estimator is too noisy; the only way to determine which is to carry 
the calculation to its conclusion and see how much reweighting increases the 
overall error bar.

There are two difficulties which make this SNR a guideline, rather than a 
quantitative measurement:
\begin{enumerate}
\item Determining the gauge variance is difficult, since it requires knowledge 
of the true weight factors, the same quantities whose estimation we are concerned 
with.
\item When using a highly diluted estimator (which we will choose to use in the 
end), determination of the stochastic variance requires computation of multiple 
stochastic estimates. This may involve a substantial amount of computer power.
\end{enumerate}
We will return to these issues later in the discussion of specific estimators 
in Sec.~\ref{sec-mapping}.

\subsection{Mapping the stochastic noise}
\label{sec-mapping}

It can be shown readily that the variance of the stochastic estimator 
$\Tr \mathcal O = \LL \xi^\dagger \mathcal O \xi \RR$ is 
\beq
\LL (\xi^\dagger \mathcal O \xi)^2 \RR - \LL \xi^\dagger \mathcal O \xi \RR^2 = 
\sum \limits_{i \neq j} |\mathcal O_{ij}|^2,
\label{eq-var}
\eeq
the sum of the squares of the off-diagonal elements of $\mathcal O$. 
Understanding which of these elements dominate is useful for designing improvements 
to the stochastic estimator. As we cannot even afford to compute all of the 
diagonal elements (to get an exact value for $\Tr \mathcal O$), we certainly cannot 
compute all of the $\mathcal O_{ij}$'s. However, we can examine a representative set 
to see which are dominant. On a single configuration from our $24^3\times 48$ ensemble, 
we have computed all $\mathcal O_{ij}$ for a set of sources $j$
\beq
{\cal S}= \left\{ j | j_{x, y, z}\in\{8, 16\} \,, j_t \in \{16, 24, 32\}\right\}\,. 
\eeq
Since we compute all spin-color combinations, the number of sources is 
$|{\cal S}|=12\times24=288$. We kept the information only for sinks $i$
such that the vector 
between $i$ and $j$ has no components larger than 12 (after accounting for periodic 
boundary conditions in the $y$ and $z$ directions). 
The number of data points is very large and to produce a more manageable set we bin
the points in equivalence classes. For the purpose of this illustration, we assume that 
all source positions are equivalent, so we average together the squares of all matrix 
elements corresponding to different source points. We notice no significant effects on 
matrix elements whose sinks are near the Dirichlet boundary, so we bin together the
points where the separation vector between $i$ and $j$ is related by a reflection in any direction
or rotation in the $(y,z)$-plane. We also bin together the points that have the same
starting and ending color indices and separately the ones that have different color 
indices. We treat the directions $x$ and $t$ separately due to the 
effects of the electric field and collect each of the 16 spinor combinations in a separate bin.
All these data points are used to create Fig.~\ref{fig-ope-map} and to predict
the error of the stochastic estimators for different dilution schemes.

The relative size of the off-diagonal elements as a function of the Euclidean separation 
between $i$ and $j$ is shown in Fig.~\ref{fig-ope-map} for 
both $M'M_0^{-1}$, the first-order term, and $M''M_0^{-1} - (M'M_0^{-1})^2$, the second
order term. We note that the magnitude of $|\mathcal O_{ij}|^2$ decreases as $i$ and $j$ are
further apart, as expected. The short-range behavior is the source of our problem.
The trace estimator we use works well for diagonally dominated
matrices, where the largest elements of the matrix lie along the diagonal, and 
decay quickly as we go away from it. Unfortunately, for our matrices the dominant
elements are not on the diagonal, as can be easily seen from Fig.~\ref{fig-ope-map}.
Even among the elements at Euclidean
separation 0, those off-diagonal elements $\mathcal O_{ij}$ where $i$ and $j$
differ in spin and color indices are larger than the diagonal elements. This is a simple 
depiction of why this stochastic estimator is so difficult: the diagonal elements 
(the signal) are small, while the near-diagonal elements contributing to noise are much larger. 
The structure is not unexpected, since $M'$ amounts to a point-split operator in the $t$ direction.

Fig.~\ref{fig-ope-map} suggests that the most direct route to reducing the variance is 
reducing the short-range off-diagonal elements of the operators. There are two somewhat 
redundant techniques we can use to do this: hopping parameter expansion improvement 
and dilution. Hopping parameter expansion has the advantage that its numerical cost is relatively
modest for small orders, but it only cancels the off-diagonal elements approximatively.
We explored this technique in a previous study using an expansion up to $7^{th}$ order,
the largest order we could afford~\cite{Freeman:2013eta}. We found that the improvement
was insufficient and the signal-to-noise ratio for polarizability was smaller than one.
In this work we explore an alternative approach: dilution.

\subsection{Dilution scheme}
\label{sec-dilution}

\begin{figure*}
  \centering\includegraphics[width=0.45\textwidth]{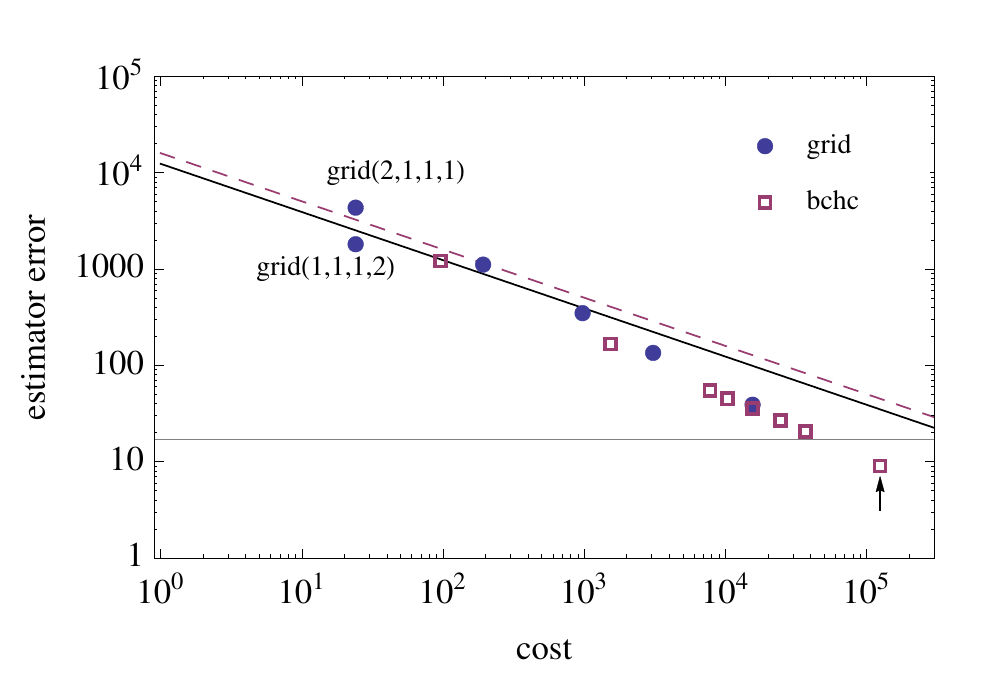}
  \hspace{0.0\textwidth}
  \includegraphics[width=0.45\textwidth]{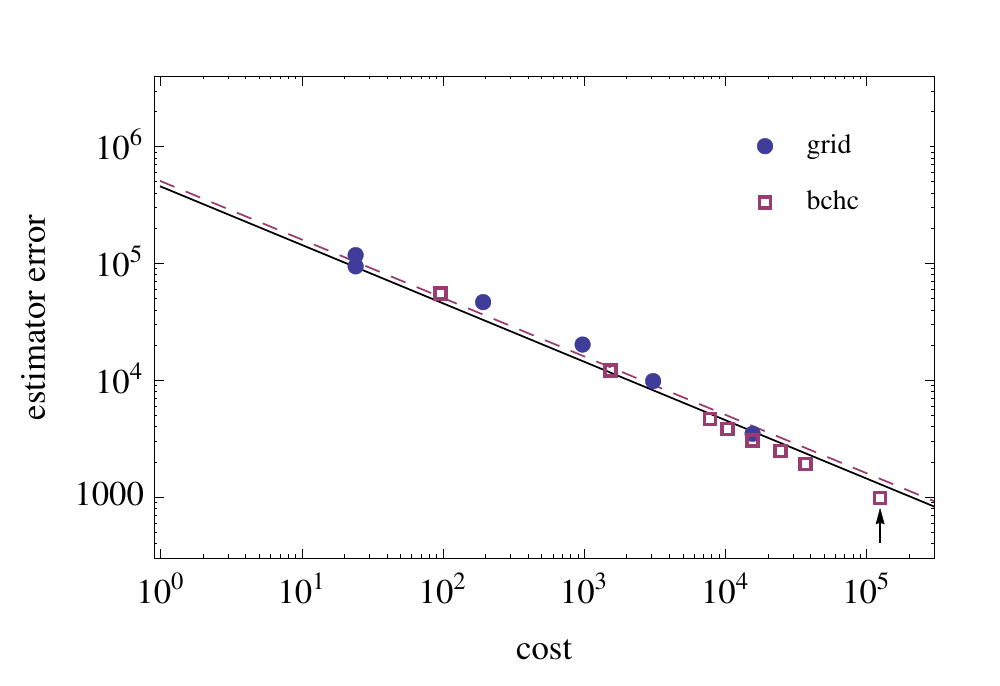}
  \caption{Uncertainty estimates for different dilution schemes for the 
  first-order term~(left) and second-order term~(right) as a function of the 
  the estimator cost, given by the number of partitions in each dilution scheme. The solid~(dashed) line indicates
  the expected uncertainty for the estimates based on repeated use of the 
  undiluted~(spin-color diluted) estimator. The horizontal line in the left panel 
  corresponds to the standard deviation of the gauge fluctuations, as estimated 
  in the next section.
  The point in the bottom-right labelled by a black arrow is the BCHC-$6^4$ dilution scheme
actually used in the computation.}
\label{fig-dilution-schemas}
\end{figure*}

Dilution is a technique which, with a suitable dilution scheme, can eliminate 
the noise contribution from near-diagonal elements. It entails partitioning the lattice 
into $N$ subspaces, estimating the trace over each separately, and adding the estimates; 
this is done in practice by generating noise vectors with support only on one subspace. 
This eliminates contributions to the variance from off-diagonal elements $\mathcal O_{ij}$ 
where $i$ and $j$ belong to different subspaces, at the cost of requiring $N$ evaluations 
of $\mathcal O$ to generate a single estimate. 
Thus there is a fundamental tradeoff involved in dilution. The aim of any stochastic 
estimation procedure is to minimize the uncertainty in the stochastic estimate for a 
given computational effort, and that uncertainty, rather than the variance of the 
estimator itself, should be used as the yardstick for measuring the utility of a 
dilution technique. 

The variance of the diluted estimator should then be compared with the variance of an 
estimate based on the average of $N$ independent evaluations of the undiluted estimator. 
The variance of this mean is smaller by a factor of $N$ than the variance of a single
evaluation. To be more precise, if we label the partition to which
the (spin/color/spatial) index $i$ belongs as $P(i)$, the variance becomes
\beq
\mathrm{var} (\Tr \mathcal O) = \sum \limits_{i \neq j} |\mathcal O_{ij}|^2 \delta_{P(i) P(j)},
\label{eq-var-dil}
\eeq
that is, the sum of only those off-diagonal elements that connect indices belonging to 
the same subspace. If all $N$ subspaces are of equal size (which is generally the case), 
then this results in a sum with only $1/N$ as many terms.
The ratio of uncertainties (the proper figure of merit) between an $N$-subspace dilution 
and the mean of $N$ undiluted estimators, is
\beq
\frac{\sum \limits_{i \neq j} |\mathcal O_{ij}|^2 \delta_{P(i) P(j)}}
{\frac1N\sum \limits_{i \neq j} |\mathcal O_{ij}|^2}.
\eeq
Thus dilution will only be a success if the average of the off-diagonal elements that 
survive (belong to the same subspace) is less than the average of all of them. Choosing 
a dilution strategy, then, must be done with consideration of the form of $\mathcal O_{ij}$, 
as it is entirely possible to partition the lattice in such a way to make the stochastic 
noise worse.

The most common sort of dilution is spin/color dilution, where each noise vector has 
support for a single spin and color over the entire lattice. As we can see from
Fig.~\ref{fig-dilution-schemas} this dilution 
scheme alone does not help us; it must be used alongside other dilution schemes in which the subspace structure 
also involves spatial separation.

To construct the spatial structure for a dilution scheme for an operator whose 
off-diagonal elements are expected to decrease with increasing Euclidean distance, 
we want to allocate sites among the $N$ subspaces so as to maximize the minimum 
Euclidean distance separating two sites belong to the same subspace. We investigate
two schemes: regular grid and body-centered hypercubic (BCHC) scheme. 

For a regular grid two points belong to the same partition if 
\beq
p_1 - p_2 = 0 \pmod \Delta \,.
\eeq 
The four-dimensional vector $\Delta$ defines the steps of the grid in the four spatial
directions. The number of partitions, which is proportional to the cost of the dilute
estimator, is controlled by the volume of one grid cell $N=\prod_i\Delta_i$. When used
in conduction with the spin-color dilution, we have $N=12\prod_i\Delta_i$. The
minimum Euclidean distance between two points on the same grid is the smallest
grid step $\min_i\Delta_i$.


In the BCHC scheme, two points belong to the same partition if
\beq
p_1 - p_2 \in \{0, \Delta\}\pmod{2\Delta}  \,.
\eeq
This can be thought of as two regular grids of steps $2\Delta$ displaced by vector $\Delta$,
so that the origin of the second grid lies in the middle of the unit cell $2\Delta$,
creating a body-centered hypercubic pattern with unit cell $2\Delta$.
The number of partitions for this scheme is $N=8\prod_i\Delta_i$, or $N=96\prod_i\Delta_i$
when spin-color dilution is also used; it is half that of a grid dilution scheme with the same
nearest-neighbor distance. The minimal
distance between two points from the same partition depends on the relative magnitude
of the components of $\Delta$; when all components are the same $\Delta_i=b$, the
minimal distance is $\left\|{\Delta}\right\|=2 b$. Note that for a regular grid we would 
need $N$ twice as large to achieve this minimal distance.

A disadvantage of large-$N$ dilution strategies is the need for a large numerical effort 
to compute even a single estimate of the trace, even if that single estimate has greatly 
reduced stochastic error. This makes it difficult to empirically determine the 
variance of a large-$N$ dilution scheme by repeated application, because the cost of 
repeating the estimator enough times to achieve a sufficiently low error on the variance
becomes prohibitive. However, we can use the off-diagonal element mapping data to 
estimate the variance for any dilution scheme. Under the assumptions outlined above,
we estimate 
\beq
\mathrm{var}(\Tr\mathcal O) \approx \frac{12\times V_\text{lat}}{N} \sum_{i \in V}^N \sum_{j \neq i}
|\mathcal O_{ij}|^2 \delta_{P(i) P(j)}\,,
\eeq
where $V_\text{lat}$ is the lattice volume and the positions and spin/color of the $N$ sources $i$ are chosen randomly.
The sum over the sinks $j$ extends over the entire lattice, rather than a limited hypercube as in Fig.~\ref{fig-ope-map},
eliminating any effects from small points beyond the horizon on the variance. Additionally, scattering the points
over the entire lattice, rather than confining them to a central region away from the Dirichlet walls, correctly
incorporates the finite-size effects from the Dirichlet boundary conditions into the estimator variance.
For $N$ large enough, the result should quickly converge to the true variance of $\mathcal O$.
Using 300 randomly chosen lattice points and evaluating all 12 color-spin indices at these points, we
determine the standard deviation for our estimators with percent-level error on a few configurations. 
We find that the standard deviation varies very little from configuration to configuration.
The mean value over the configurations is used for the data in Fig.~\ref{fig-dilution-schemas}. 


This is a useful tool to use in planning a dilution scheme. In 
Fig.~\ref{fig-dilution-schemas} we compare the predicted uncertainty for our stochastic
estimator using various dilution schemes. Except for the solid line, all estimators
use spin-color dilution. As noted before, the spin-color dilution by itself 
(indicated by the dashed line) is inferior to the undiluted estimator. At first order, moderately-aggressive
dilution schemes essentially keep pace with the decline in the estimator variance caused by
simple repetition. Dilution begins to win out once the minimum Euclidean distance between
adjacent points in the same subspace, reflected in the increasing cost, increases. At second order,
only an extremely small improvement is seen; this is due to the substantially slower falloff
of the offdiagonal elements seen in Fig.~\ref{fig-ope-map}. Either a more aggressive dilution scheme
or an operator-improvement technique used in tandem with dilution is needed to see much improvement 
over simple repetition of the na\"{i}ve estimator.

The BCHC dilution schemes should show at best a reduction in the cost by a factor of $2$
compared to grid schemes, since they achieve the same minimum distance
with half as many partitions. The actual gain is less than this, because a grid source has
only eight nearest neighbors, while the BCHC source has sixteen. Nonetheless, for both the 
first and second order estimators, the BCHC dilution outperforms grid dilution by a small
amount.

To reduce the 
stochastic variance to a level comparable with the gauge variance we need a large grid spacing.
In the left panel of Fig.~\ref{fig-dilution-schemas} we see that this happens for the
first-order derivative only when $\Delta = \{6,6,6,6\}$. This is the dilution scheme used in
the subsequent calculation.
In this scheme, the minimal Euclidean distance between two points in the same partition is
$12$ and the number of partitions is $N=96\prod_i\Delta_i=124,416$.

\subsection{Gauge variance}

Off-diagonal element data allows us to determine the expected variance for our estimators.
However, it provides no indication as to the level of gauge variance, which we also need 
to know to determine whether a dilution scheme noise is smaller that the expected signal,
as discussed in Section~\ref{sec-est-qual}. To estimate gauge variance we did two tests:
an extrapolation from small lattices where we can compute the operators exactly and
a more computational intensive study where we evaluated our expensive high quality 
estimator (BCHC with $\Delta = \{6,6,6,6\}$) on a couple of lattices from our ensemble.

\begin{table}[t]
\begin{tabular}{@{}ccrrlllr@{}}
\toprule
\multirow{2}{*}{config} & \multirow{2}{*}{$N_\text{est}$} & \phantom{a} &\multicolumn{2}{c}{$w_q'$} &\phantom{a}& \multicolumn{2}{c}{$\tilde{w}_q''$}\\ 
\cmidrule{4-5} \cmidrule{7-8} 
                        &                    && \multicolumn{1}{c}{mean} & \multicolumn{1}{c}{std-dev} && \multicolumn{1}{c}{mean} & \multicolumn{1}{c}{std-dev} \\ \midrule
2                       & 6                  && -2.8(2.7)                 & 6.5(2.1)                     && -196,362(468)             & 1147(371)                     \\ 
3                       & 4                  && -19.9(4.7)                & 9.4(4.0)                     && -197,399(324)             & 648(274)                     \\ 
\bottomrule
\end{tabular}
\caption{Repeated trials of the BCHC diluted estimator for two configurations. 
The standard deviation field indicates the stochastic error, which we determined in 
Section~\ref{sec-dilution} to be $9.5(6)$ for the first-order estimator
and $1038(76)$ for the second-order one.}
\label{table-test}
\end{table}

We discuss first direct evaluation of our estimator on $24^3\times 48$ lattices. For the
first two lattice configurations in our ensemble we run several evaluations of our estimator.
The results of this test are shown in Table \ref{table-test}.
We first note that the standard deviation for the stochastic estimators is consistent
with the estimate from the previous section.
For the first-order term $w_q' = \Tr M' M_0^{-1}$ the gauge fluctuations 
are $16(8)$. This estimate takes into account the fact the gauge average is zero, by 
reflection symmetry, for the first-order term. A correction factor is used to account
for the bias in the standard deviation estimator. The stochastic fluctuations are 
smaller than the gauge fluctuations. This suggests that this estimator 
is precise enough to follow the gauge fluctuations.

For the second-order term, the gauge average value is $-196,881(491)$. 
The standard deviation of the gauge fluctuations is 
$\sigma_\text{gauge}=919(694)$, of similar order
with the stochastic uncertainty. It is not clear whether the signal-to-noise
ratio is good enough for this estimator, especially since our determination
is also compatible with small values for the gauge fluctuations. We will see 
that the extrapolation from small volumes predicts that~$\sigma_\text{gauge}$ 
is on the small side of the estimate. This suggests that the second order estimator
is noisy. Note that the cost of this seemingly-simple study is 2.5 million 
inversions, about $3\%$ of the cost of the entire calculation.

We turn now to the extrapolation from small volumes. We generated a set of small
lattice of different geometries and computed the first and second order derivatives
exactly using the compression method for Wilson 
fermions~\cite{Alexandru:2010yb,Nagata:2010xi}. More precisely, we computed the 
fermionic determinant on these lattices exactly for $7$ different
values of the electric field parameter $\eta$ and then evaluated the derivatives
numerically using a $\mathcal O(\eta^6)$ finite difference scheme 
\beq
f' \approx \frac1\eta\sum_{k=-3}^3 c'_{k} f(k\eta) \,,\kern0.5em
f'' \approx \frac1{\eta^2}\sum_{k=-3}^3 c''_{k} f(k\eta) \,,
\eeq
where $f(\eta)=\log\det M(\eta)$. It is straightforward to relate these derivatives
to the derivatives of the reweighting factor: $w'_q=f'$ and $\tilde{w}''_q=f''$.
The coefficients for these approximations are given in Table~\ref{tab-coef-deriv}.
We use a value of $\eta=0.01$ which is sufficiently precise.

\begin{table}[t]
\begin{tabular*}{0.9\columnwidth}{@{\extracolsep{\stretch{1}}}*{9}{r}@{}}
\toprule
$k$   &\phantom{ab}& -3    & -2    & -1   & 0      & 1   & 2       & 3    \\
\midrule
$c_k'$  && -1/60 & 3/20  & -3/4 & 0      & 3/4 & -3/20 & 1/60 \\
$c''_k$ && 1/90  & -3/20 & 3/2  & -49/18 & 3/2 & -3/20   & 1/90 \\
\bottomrule
\end{tabular*}
\caption{Coefficients for the finite difference derivatives.}
\label{tab-coef-deriv}
\end{table}

For each lattice geometry we generated 10 configurations. We used Wilson pure gauge
action with $\beta=6.0$. The lattice spacing is $a/r_0 = 0.186$~\cite{Necco:2001xg}, 
which is similar to the lattice spacing for our large configurations. For the fermionic 
matrix, we use nHYP fermions with $\kappa=0.1267$. The parameter $\kappa$ was 
adjusted to produce a pion mass around $300\MeV$ to match the sea quark mass on 
the large configurations.

To make sure that we are not in the deconfined phase, we have to keep 
$r_0/L < r_0 T_c = 0.7498(50)$~\cite{Necco:2003vh}. This means that all of
our lattice dimensions $n_i=L_i/a$ should satisfy $n_i \geq 8$. Since this 
is already at the upper range of lattice volumes where we can compute the 
determinant exactly, to investigate a wider range
of volumes we have to use geometries that do not satisfy this constraint. For
these lattices, we take advantage of the Dirichlet boundary conditions in the
$x$ and $t$ directions and cut out these lattices from larger ones, with $n_x=n_t=12$, that 
are in the confined phase. The only delicate step in this process is
that we have to smear the links on the larger lattice and then cut it, 
so that the boundary do not introduce discontinuities. We use 72 different 
lattice geometries: $n_y,n_z\in\{8,10,12\}$, 
$(n_x|n_t)\in\{4|4, 4|6, 6|4, 4|8, 8|4, 4|10, 10|4,6|6\}$.

\begin{figure*}[t]
\includegraphics[width=1\textwidth]{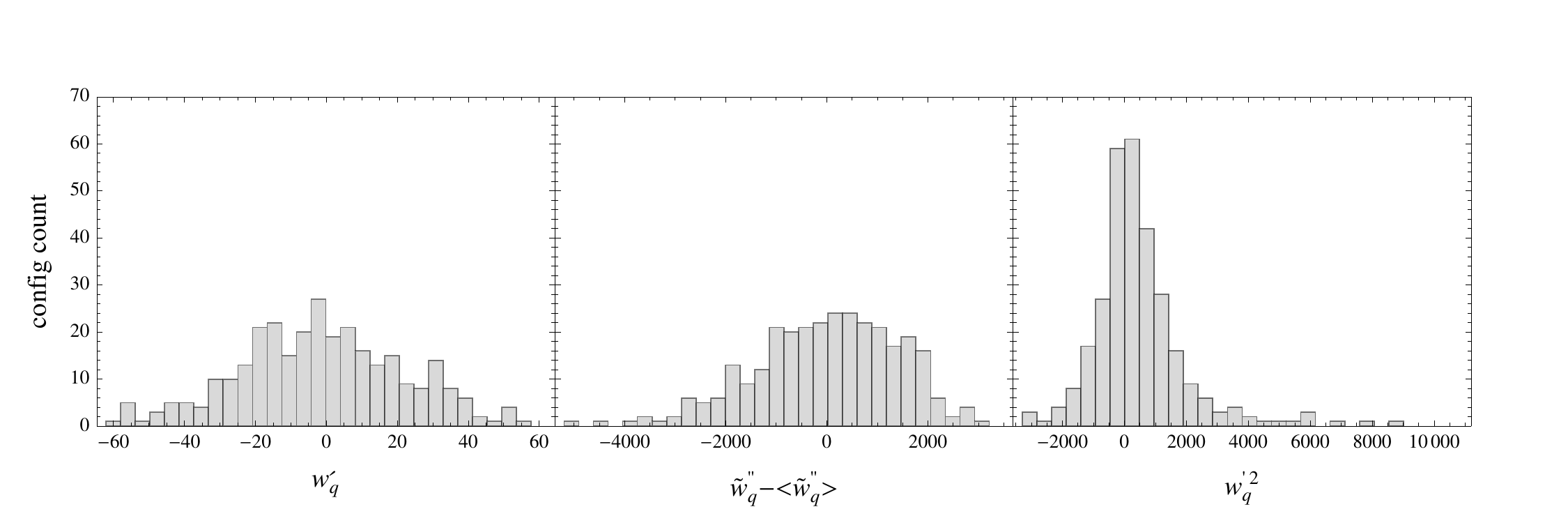}
\caption{Distribution of the values of the stochastic estimators on the full 
300-configuration ensemble. The two leftmost panels are the stochastic estimators
for $w_q'$ and $\tilde{w}''_q$ using BCHC dilution. The right panel is the
estimator for $w_q'^2$ using a combination of BCHC-dilution and hopping parameter
expansion estimators, as described in the text. Note that for the center panel
we shifted the distribution by subtracting its mean.}
\label{wf-fig}
\end{figure*}

For each ensemble we determine the gauge standard deviation for both derivatives 
and mean for the second derivative. We analyzed the dependence of each of these 
three quantities as we varied the dimension of the lattice in each direction. In
most cases we found that these quantities vary linearly with the dimension (either
relatively constant or raising linearly). The only exception is the mean of the
second order derivative which requires quadratic terms to describe its dependence
on $n_x$, the extent of the lattice in the direction of the external field.
Based on these observations and taking into account the rotational symmetry in the 
$(y,z)$-plane, the fit functions we use in our extrapolations are
\beqs
\sigma_{w_q'} &= \alpha (n_x + \beta_x)(1+ \gamma n_y)(1+ \gamma n_z)(n_t + \beta_t)\,, \\
\av{\tilde{w}_q''} &= \alpha (n_x + \beta_x + \gamma n_x^2) n_y n_z (n_t + \beta_t) \,, \\
\sigma_{\tilde{w}_q''} &= \alpha (n_x + \beta_x)(1 + \gamma n_y)(1 + \gamma n_z)(n_t + \beta_t)\,.
\eeqs
The results of the fits are presented in Table~\ref{tab-fit-coef}. Using these coefficients
and their cross-correlations, we estimate that for a $24^3\times 48$ lattice the gauge averages
and standard deviations should be
\beqs
\sigma_{w'_q} &= 17(4)\,,\\
\av{\tilde{w}''_q} &= -212(2)\times10^3\,,\\
\sigma_{\tilde{w}_q''} &=164(62)\,.
\eeqs
We note that all these results are compatible with the values determined via repeated
evaluation of the stochastic estimator on two full-size configurations. As we mentioned
earlier, the gauge standard deviation for $\tilde{w}''_q$ is lower than the stochastic 
uncertainty, indicating a noisy estimator.

\begin{table}[b]
\begin{tabular*}{0.9\columnwidth}{@{\extracolsep{\stretch{1}}}*{8}{r}@{}}
\toprule
            &\phantom{ab}&  Q    & $\alpha$    & $\beta_x$& $\beta_t$& $\gamma$  \\
\midrule
$\sigma_{w_q'}$          && 0.15 &  0.0017(10) & -1.6(3)  & 1.9(1.2) & 0.08(4)  \\
$\av{\tilde{w}_q''}$     && 0.85 & -0.09(2)    & -3.95(3) & -1.61(7) & 0.118(5)  \\
$\sigma_{\tilde{w}_q''}$ && 0.27 &  0.09(3)    & -3.1(1)  & 3(2)     & 0.014(16)\\
\bottomrule
\end{tabular*}
\caption{Fit parameters for extrapolation from small volumes. Q is the confidence
level of the fit.}
\label{tab-fit-coef}
\end{table}

\section{Results}
\label{sec:results}

\subsection{Reweighting factors}

Before we turn to the main results in this paper, hadron polarizabilities, we present
the results for the reweighting factors, as evaluated on the full ensembles using
the estimators described in the previous section.

The resulting estimates for $w'_q$, $\tilde{w}''_q$, and $(\Tr M' M_0^{-1})^2$ are given 
in Fig. \ref{wf-fig}. We discuss here briefly the estimator for 
$w_q'^2=(\Tr M' M_0^{-1})^2$. When more than one estimate per configuration of the 
first-order term $w_q'$ is available, such as in the previous study using hoping parameter 
expansion improvement where we used thousands of cheap estimators per 
configuration~\cite{Freeman:2013eta}, we 
may construct one estimate for $w_q'^2$ out of two independent estimates of $w_q'$. 
However, in this study we used an expensive BCHC-diluted estimator and there is no second 
estimate of $w_q'$ available. Constructing a second one in the same manner as the first, 
using the $N=124,416$ dilution scheme, would require a large extra effort. However, we 
observed from the previous study that the stochastic fluctuations of this term compared 
to the fluctuations of the rest of the traces involved in $\tilde{w}_q''$ are small. Thus 
it is acceptable to use a less labor-intensive method to estimate it. Since we have the 
estimates of $w_q'$ from the prior run saved to disk, we use them in combination with the new 
diluted estimates of $w'_q$ to produce an estimate of $w_q'^2$ on each configuration. 

For the first order term we find that the standard deviation is $\sigma_{w_q'}=23(1)$.
This includes both the stochastic noise and the gauge fluctuations. The determination is
compatible with our estimates described in the previous section.
For the second order term, $\tilde{w}_q''$, the mean value is $\av{\tilde{w}_q''}=-197,549(83)$
and the standard deviation is $\sigma_{\tilde{w}_q''}=1429(58)$, again in agreement
with the values estimated in the previous section. We note that the combined standard deviation
is larger than the gauge one estimated from the extrapolation from small volumes, indicating 
that the stochastic noise is dominant for this estimator.

For the $w_q'^2$ estimator we find that the standard deviation is $\sigma_{w_q'^2}=1655(68)$.
This is comparable with the standard deviation for $\tilde{w}_q''$ and it would seem that
this term will add significant variance to the final result. To see why this term is subdominant
we have to expand Eq.~\ref{eq-expansion} in terms of traces:
\beq
w(\eta_d) = 1-\eta_d w_q' + \frac12\eta_d^2 \left( 5\tilde{w}_q'' + w_q'^2 \right) \,.
\eeq
We see that in the final result, the quadratic term $w''=5\tilde{w}_q'' + w_q'^2$
is dominated by $\tilde{w}_q''$. Indeed the total standard deviation for the quadratic term 
is $\sigma_{w''}=7231(295)$, compared to the contribution coming from $\tilde{w}_q''$ alone, 
$5\times\sigma_{\tilde{w}_q''}=7147(292)$.

\begin{table*}[t]
\begin{tabular*}{0.95\textwidth}{@{\extracolsep{\stretch{1}}}l*{16}{r}@{}}
\toprule
        &\phantom{ab}& \multicolumn{3}{c}{Valence only} &\phantom{a}& 
        \multicolumn{3}{c}{$1^\text{st}$ order} &\phantom{a}& \multicolumn{3}{c}{$\tilde{w}_q''$ only} 
        &\phantom{a}& \multicolumn{3}{c}{$2^\text{nd}$ order} \\
\cmidrule{3-5}\cmidrule{7-9}\cmidrule{11-13}\cmidrule{15-17} 
        && \multicolumn{1}{c}{$aE$}          & \multicolumn{1}{c}{$a\Delta E$}          & \multicolumn{1}{c}{$Q$} 
        && \multicolumn{1}{c}{$aE$}            & \multicolumn{1}{c}{$a\Delta E$}             & \multicolumn{1}{c}{$Q$} 
        && \multicolumn{1}{c}{$aE$}            & \multicolumn{1}{c}{$a\Delta E$}              & \multicolumn{1}{c}{$Q$} 
        && \multicolumn{1}{c}{$aE$}            & \multicolumn{1}{c}{$a\Delta E$}              & \multicolumn{1}{c}{$Q$}       \\
\midrule
Pion    && 0.245(1)   & -5.4(3.4)   & 0.17  && 0.245(1)     & -6.0(3.4)     & 0.18     && 0.245(1)     & 5.4(5.6)       & 0.15     && 0.245(1)     & 5.6(5.7)       & 0.15    \\
Kaon    && 0.352(1)   & 4.2(0.8)    & 0.12  && 0.352(1)     & 3.7(1.0)      & 0.07     && 0.352(1)     & 10.5(3.4)      & 0.03     && 0.352(1)     & 11.1(3.4)      & 0.02    \\
Neutron && 0.694(4)   & 62.8(5.7)   & 0.65  && 0.694(4)     & 63.9(6.5)     & 0.57     && 0.695(4)     & 72.5(16.4)     & 0.53     && 0.695(4)     & 67.0(16.3)     & 0.43    \\
\bottomrule
\end{tabular*}
\caption{Results for the energy and energy shift for the pion, neutron, and kaon with differing orders 
of reweighting: none (the valence-only calculation), first-order in $\eta_d$, second-order 
including only the dominant contribution in $\tilde{w}_q''$, and the full calculation to 
second order. For the energy shifts, the values are in units of $10^{-8}$. $Q$ is the confidence
level for the fits corrected to account for the sample size~\cite{Toussaint:2008ke}.}
\label{table-results}
\end{table*}

\subsection{Hadron polarizabilities}

The power series expansion given in Eq. \ref{eq-expansion} can be used to determine the 
weight factor at any desired $\eta_d$ on each configuration; these weight factors can 
then be combined with the valence correlators computed previously to complete the 
calculation. We note that one set of weight factor estimates may be used without 
modification for all hadrons; this is a strength of the reweighting approach. Full 
details of the valence correlators are given in~\cite{Lujan:2014kia}; we repeat only the 
essential elements here. We use point interpolators for both source and sink. To improve 
the signal-to-noise ratio, we use 28 sources per 
configuration; in any case the expense of the many sources is dwarfed by the cost of the 
weight factor estimates. These sources are spread evenly in the $(y,z)$-plane but are 
along the centerline $x=12$ to avoid the Dirichlet walls.

It is informative to turn on the reweighting one order at a time; we additionally add 
the extra second-order term, $w_q'^2$, separate from the others 
that comprise $\tilde{w}_q''$. Using these approximations for the reweighting factors
we compute the hadron propagators using Eq.~\ref{eq:rew} and do a correlated fit
for zero field and non-zero field propagators using the model in Eq.~\ref{eq:model}.
The fit ranges for these fits are the same as in our previous valence study~\cite{Lujan:2014kia}.
The results for these fits are presented in Table \ref{table-results}.
Focusing on the energy shift, $a\Delta E$,
note that the uncertainty remains relatively constant when including only the first
order terms, indicating that our estimator adds very little noise. The second order
term, in particular $\tilde{w}_q''$, introduces significantly more uncertainty, 
doubling or trebling the size of the error bars. 
In principle this could be due to either the gauge fluctuations of the second-order 
term causing a large fluctuation in the weight factor. However, in our case the estimated 
stochastic error for $\tilde{w}_q''$ is fairly
large compared to the overall variation of the estimator, so we suspect that the largest 
share of the fluctuations in our estimates are due to stochastic noise, despite the 
substantial effort involved in the estimator.
As discussed previously, the
addition of the $w_q'^2$ estimate has very little effect both on the value of the
energy shift and its error.

To convert the energy shift to polarizability we use the relation:
\beq
\alpha = \frac{2a^3 e^2}{9\eta_d^2} (a\Delta m) = \frac{2a^3 e^2}{9\eta_d^2} \frac{a E}{a M} (a\Delta E)\,,
\eeq
where $M$ is the mass of the hadron of interest, computed using periodic boundary conditions. These
masses were computed for this ensemble in a previous study~\cite{Lujan:2014kia}.
For the neutron a correction due to the magnetic moment is required, 
$\alpha_c=\alpha+\mu^2/(2m)$~\cite{Lvov:1993fp,Detmold:2010ts,Lujan:2014kia}.
The polarizability values are given in Table~\ref{table-results-pol}. We
will discuss now each hadron separately.

\begin{table}[b]
\begin{tabular*}{\columnwidth}{@{\extracolsep{\stretch{1}}}lcrrrr@{}}
\toprule
&\phantom{ab}& Valence only & $1^{\rm {St.}}$ order & $\tilde{w}_q''$ only & $2^{\rm {nd}}$ order \\
\midrule
Pion && -0.21(14) & -0.24(14) & 0.21(22) & 0.22(23)\\
Kaon && 0.14(3) & 0.13(3) & 0.36(12) & 0.38(12)\\
Neutron &&2.56(19) & 2.60(22) & 2.89(55) & 2.70(55)\\
\bottomrule
\end{tabular*}
\caption{Electric polarizability for the pion, neutron, and kaon with differing orders 
of reweighting: none (the valence-only calculation), first-order in $\eta_d$, second-order 
including only the dominant contribution in $\tilde{w}_q''$, and the full calculation to 
second order. Values are in units of $10^{-4}\fm^3$.}
\label{table-results-pol}
\end{table}

We remind the reader that the neutral pion correlator used in this study does not include
the disconnected diagrams that are required due to the isospin breaking introduced by the
electric field. This is a common limitation for lattice calculations, since the inclusion of
these terms is computationally expensive. For the neutral pion, chiral perturbation theory
predicts a polarizability around $\alpha_{\pi^0}\approx-0.5\times10^{-4}\fm^3$. In the absence 
of the disconnected contributions, the prediction is that the polarizability would be positive 
and an order of magnitude smaller in absolute value~\cite{Detmold:2009dx}. Lattice calculations 
of this quantity indicate that the connected neutral pion polarizability turns negative as the 
pion mass is lowered below~$400\MeV$, contradicting these 
expectations~\cite{Detmold:2009dx,Alexandru:2010dx,Lujan:2014kia}.
It was suggested that this discrepancy is due to final volume corrections~\cite{Detmold:2009dx}, but
this does not seem to be the case~\cite{Alexandru:2010dx}. The correction associated with charging
the sea quarks might also be responsible for this discrepancy. As we can see from Table~\ref{table-results-pol},
the polarizability for the neutral pion seems to change signs as we charge the sea quarks. However,
the current errors are too large, relative to the size of polarizability, so no definitive conclusions can be
drawn. We also measured the change in the energy shift induced by the reweighting taking into account the
correlations between the original and reweighted measurements; the result was consistent with zero.
We note that the valence-only result is very close to zero; the ensemble in question happens to lie 
very near the value of $m_\pi$ where the polarizability of the neutral pion changes 
sign~\cite{Alexandru:2010dx,Lujan:2014kia}. 

Neutral kaon polarizability is not shifted by the first order reweighting. When the second
order is included, both the central value and its uncertainty increase. In this case the shift
in polarizability is statistically significant. It is interesting to note that this behavior is
consistent with the features we observed in our previous study: kaon polarizability was
insensitive to the change in mass of the valence light quarks, but it shifted significantly
when the mass of the light sea quarks was changed~\cite{Lujan:2014kia}. This was in contrast
with the pion polarizability which seems to depend strongly on the valence quark mass, but it
was fairly insensitive to the sea. It is then not surprising that the kaon polarizability should 
be sensitive to charging the light sea quarks. In any case, the chiral extrapolation performed
in our previous study for kaon polarizability needs to be revisited, given the significant shift 
induced by charging the sea.

The neutron, the benchmark hadron for this type of calculations, shows no statistically significant change when the 
coupling to the sea is turned on via reweighting. This is a bit puzzling since the chiral perturbation
theory expectation is that the neutron polarizability increases by $1$--$2\times10^{-4}\fm^3$ when
the sea quarks are charged~\cite{Detmold:2006vu} and our errorbars, even after including the second order
correction, are small enough to resolve this difference. It is still possible that this increase shows
up after removing the finite volume effects, that are expected to be significant for this quantity.
We note that our calculation of neutron polarizability, including sea effects, improves upon the 
precision of the only other such calculation known to us~\cite{Engelhardt:2007ub,Engelhardt:2010tm} 
in both precision and pion mass.

While the effects of charging the sea quarks are not statistically significant here, with 
the exception of the kaon, we expect them to be enhanced both by enlarging the lattice volume 
and by approaching the chiral limit. Considering the chiral limit: when the pion mass is reduced 
it is easier to create virtual pion loops which increases the size of the pion cloud and its
contribution to polarizability. Similarly, increasing the size of the box reduces the 
momentum of the lowest pion state (recall that we use Dirichlet boundary conditions), reducing
the cost of exciting pions, with similar consequences. We thus expect the effect of charging 
the sea to be substantially larger at lower pion mass and on larger boxes.

\section{Conclusion}

While the result for the neutron here is physically significant, as it improves on the 
previously-attained precision, we treat it more of a proof of concept for the perturbative 
reweighting method which will soon be applied to ensembles with larger volumes and smaller pion masses, 
where we expect the effect to be larger. The perturbative estimate for the weight factor correctly
predicts the slope of the exact determinant ratio on small lattices where it can be computed exactly, 
but like the conventional reweighting estimator it is quite noisy. However, dilution can be used to reduce 
its variance. Strong dilution with the body-centered hypercubic pattern outperforms
hopping parameter expansion and it is certainly simpler to formulate and more flexible.

Our results suggest that while these estimates of the first-order term $w'_q$ are sufficient, 
a reduction in the stochastic noise from the second-order term would be welcome, given that the other ensembles
in the study will be inherently more expensive. Dilution completely eliminates the near-diagonal
contributions, at the cost of indirectly increasing the contributions away from the diagonal 
since we no longer average together many estimates. The long-distance behavior of the off-diagonal 
elements is exponential and its slope is governed by $m_\pi$. We are exploring the use of low-mode 
subtraction to eliminate the lowest lying modes of the Dirac operator from the operators in question 
and thus increase the exponent of the falloff; preliminary studies of this technique look promising. 

\acknowledgements

We would like to thank Craig Pelissier for generating the gauge ensemble used in this work. 
The computations were done in part on the IMPACT GPU cluster and Colonial One at GWU, the GPU cluster
at Fermilab.  This work is supported in part by the NSF CAREER grant PHY-1151648 and 
the U.S. Department of Energy grant DE-FG02-95ER40907.

\bibliographystyle{JHEP}
\bibliography{my-references}

\end{document}